\documentclass[icps3]{svjour}
\usepackage{graphics}
\begin{document}
\setlength{\topmargin}{-0.66in}
\title{Effects of a parallel magnetic field on the novel metallic behavior in
two dimensions}
\titlerunning{Effects of a parallel magnetic field on the novel metallic
behavior\ldots}
% The running title should be in less than 70 characters.
%
\author{K. Eng\inst{1,2} \and X. G. Feng\inst{1} \and Dragana 
Popovi\'{c}\inst{1} \and S. Washburn\inst{2}% etc
}                     % Do not remove
%
% Insert author list for the running head here
\authorrunning{K. Eng et al.}
% If the number of the authors is more than 3, only the first author
% should be listed and the others represented as et al.
%
\institute{National High Magnetic Field Laboratory, Florida State University,
Tallahassee, FL 32310, USA \and Department of Physics and Astronomy, 
University of North Carolina at Chapel Hill, Chapel Hill, NC 27599, USA}
\maketitle
\begin{abstract}
Magnetoconductance (MC) in a parallel magnetic field $B$ has been measured in 
a two-dimensional electron system in Si, in the regime where the conductivity 
decreases as $\sigma (n_s,T,B=0)=\sigma (n_s,T=0) +A(n_s)\,T^2$ ($n_s$ -- 
carrier density) to a {\em non-zero} value as temperature $T\rightarrow 0$.  
Very near the $B=0$ metal-insulator transition, there is a large initial drop 
in $\sigma$ with increasing $B$, followed by a much weaker $\sigma (B)$.  At 
higher $n_s$, the initial drop of MC is less pronounced.
\end{abstract}
\section{Introduction}
\label{intro}
\begin{figure}
\resizebox{0.49\textwidth}{!}{%
  \includegraphics{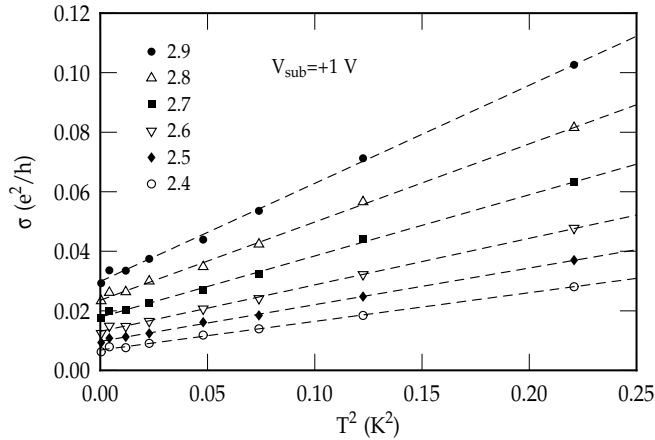}
}
%\vspace{5cm}       % Give the correct figure height in cm
\caption{Sample 19: $\sigma (T)$ in zero magnetic field plotted {\it vs.} 
$T^2$ for different $n_s (10^{11}$cm$^{-2}$) given on the graph.  The dashed 
lines are fits.  The data are shown for $0.020\leq T <0.5$~K.}
\label{st}  
\end{figure}

We have recently reported~\cite{Feng_PRL} the observation of a novel 
two-dimensional (2D) metallic behavior in Si metal-oxide-semiconductor 
field-effect transistors (MOSFETs).  In this regime, the conductivity decreases
as $\sigma (n_s,T)=\sigma (n_s,T=0)+A(n_s)\,T^2$ ($n_s$ -- carrier density) to
a {\em non-zero} value as temperature $T\rightarrow 0$.  This simple $\sigma 
(T)$ spans two decades in $T$ ($0.020 \leq T < 2$~K).  Several samples have 
been studied in detail and they all exhibit qualitatively the same behavior.
Fig.~\ref{st} shows some typical results obtained in the lowest $T$ range.  The
substrate bias $V_{sub}=+1$~V was used to maximize the range of $T$ where this
novel metallic behavior is observed~\cite{Feng_PRL}.  

In 2D, the existence of a metal with $d\sigma/dT >0$ is very surprising as it
contradicts any theoretical description available to date.  All the other
related research efforts in recent years have considered only the metallic
behavior with $d\sigma/dT <0$.  However, some of the tremendous amount of work
that was done on Si MOSFETs in earlier decades has been largely ignored.  In
particular, all the samples from previous studies could be divided into two
groups: one group exhibited the behavior consistent with that seen in 
high-mobility Si MOSFETs~\cite{Krav}, but the other displayed various
``anomalies''~\cite{AFS} that were never understood.  Our samples are
representative of the latter, a {\em broad} class of Si MOSFETs historically 
known as ``nonideal'' samples~\cite{AFS}.  We have established~\cite{Feng_PRL}
the precise form of $\sigma (T)$ in these devices.  The data strongly 
suggest~\cite{Feng_PRL} the existence of a metallic phase at $T=0$ and of a
novel, {\em continuous} metal-insulator transition (MIT) in 2D.  New 
analysis of some of the early data~\cite{Sjostrand} has 
established~\cite{kirstin} that the same $\sigma (T)$ is exhibited by a 
variety of both n-type and p-type Si inversion layers, in both circular and 
linear geometries, with different channel lengths, substrate dopings, and 
oxide thickness.  
\section{Results and Discussion}
\label{results}
Our samples are standard Si MOSFETs of Corbino geometry (channel length 
$=0.4$~mm, mean circumference $=8$~mm), and a peak mobility $\sim 1$m$^2$/Vs 
at 4.2~K.  Other sample details have been given 
elsewhere~\cite{DP_PRL,Feng_PRL}.  Conductance was measured as a function of
gate voltage $V_g$ using a low-noise current preamplifier and a low-noise 
analog lock-in at $\sim 17$~Hz.  The excitation voltage was kept low enough to
avoid electron heating.  Measurements were carried out in either a $He^3$ 
cryostat or a $He^3 - He^4$ dilution refrigerator.  $\sigma (T,n_s)$ was 
measured in fields of up to 18~T and for $0.020\leq T\leq 4$~K.  

\begin{figure}
\resizebox{0.49\textwidth}{!}{%
  \includegraphics{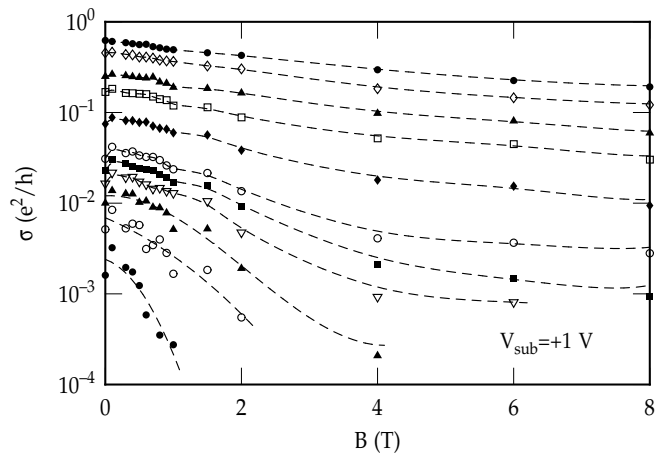}
}
%\vspace{5cm}       % Give the correct figure height in cm
\caption{Sample 9: $\sigma$ {\it vs.} $B$ applied parallel to the 2D plane at 
$T=0.25$~K for $n_s (10^{11}$cm$^{-2})=3.0, 2.8, 2.5, 2.3, 2.0$ and $1.7$ to 
$1.2$ in steps of $0.1$ (top to bottom).  At $B=0$, $n_c=0.95\times 
10^{11}$cm$^{-2}$.  Dashed lines are guides to the eye.}
\label{MC}  
\end{figure}

\begin{figure}
\resizebox{0.49\textwidth}{!}{%
  \includegraphics{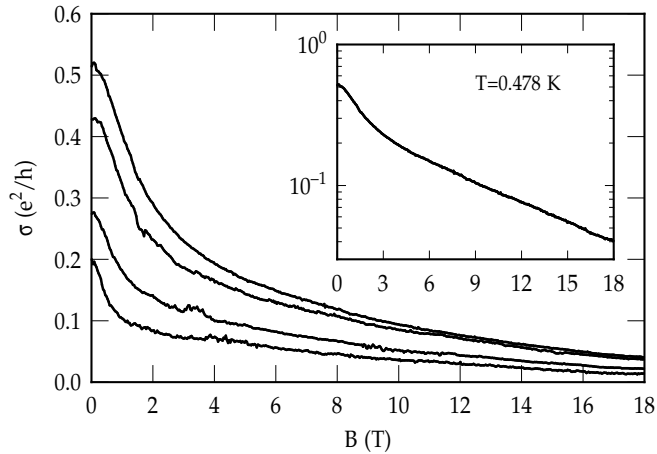}
}
%\vspace{5cm}       % Give the correct figure height in cm
\caption{Sample 19: $\sigma$ {\it vs.} parallel magnetic field $B$ for
$n_s=4.0\times 10^{11}$cm$^{-2}$ and $T=0.478, 0.403, 0.269, 0.152$~K (top to
bottom.); $V_{sub}=+1$~V.  Inset: $\sigma (B)$ at $T=0.478$~K on a semi-log
scale.  $\sigma (B)$ is exponential at the highest fields.}
\label{mkrun}      
\end{figure}

While a detailed study of
$\sigma (T,n_s,B)$ will be presented elsewhere~\cite{Eng}, here we show some
typical magnetoconductance data obtained at $T=0.25$~K for several $n_s$ above
the zero-field $n_c$ (Fig.~\ref{MC}).
Very near the $B=0$ MIT, a large initial drop of MC is observed with 
increasing $B$, followed by a weaker dependence at higher fields.  At higher 
$n_s$, the initial drop of MC appears to be less pronounced although $\sigma$ 
continues to decrease with $B$ at all temperatures (Fig.~\ref{mkrun}).  In 
fact, a closer inspection of the data (Fig.~\ref{mkrun} inset) reveals that, in
the high field regime (here $B>5$~T), $\sigma$ decreases {\em exponentially} 
with $B$.  Such a strong $\sigma (B)$ in the high field regime has been, so 
far, observed only in the very dilute 2D hole system in GaAs~\cite{Yoon} with 
the ``conventional'' metallic conduction ($d\sigma/dT <0$).  The effect has
been attributed~\cite{Sarma} to the coupling of the parallel field to the 
orbital motion arising from the finite 2D layer thickness.  In Si-based 
devices, on the other hand, a saturation in conductivity has been 
observed~\cite{par_B,Okamoto} at high fields.  Therefore, the behavior of our
samples at high $B$ differs from that of other Si MOSFETs.  For our 
experimental conditions, the thickness of the 2D inversion layer is less than 
5~nm~\cite{Kunze} (the magnetic length varies from 12~nm to 6~nm for 
$5<B<18$~T), {\it i.~e.} comparable to that in other Si MOSFETs~\cite{par_B}.
Thus it seems unlikely that the difference in the high field response can be 
explained by the orbital effects~\cite{Sarma}.  

In the low field regime, all devices seem to exhibit similar behavior.  It has
been argued~\cite{Okamoto,Vitkalov} that the magnetic field above which the 
conductivity ``saturates'' is the field required to fully polarize the spins 
of free 2D carriers.  Other experiments, on the other hand, seem to 
support~\cite{Pudalov_loc} an alternative possibility in which the parallel 
field polarizes spins in an ``external'' band of localized states, causing a 
change in scattering.  So far, however, there has been no experimental attempt
to identify the origin of this ``external'' band and to vary its population in
samples with $d\sigma/dT <0$.  In our samples, on the other hand, we have 
established~\cite{Feng_PRL} that the $T^2$ form of $\sigma (T)$ in the 
metallic regime is related to the population of localized states in the tail 
of the upper 2D subband.  The number of occupied ``upper tail'' states was 
varied systematically by applying $V_{sub}$.  It is intriguing that the same
effects might be responsible for the behavior of magnetoconductance in these 
two, apparently very different, conduction regimes (with $d\sigma/dT < 0$ and
$d\sigma/dT >0$, respectively).

While all experiments seem to suggest that spin degrees of freedom are playing
an important role in dilute 2D systems, further work is clearly needed in 
order to reach a detailed microscopic understanding of the 2D metallic phase.
In particular, ``non-ideal'' Si MOSFETs are proving to be an invaluable 
resource.

This work was supported by the National High Magnetic Field Laboratory (NHMFL)
through NSF Cooperative Agreement DMR-9527035, an NHMFL In-House Research 
Program grant, and NSF Grants DMR-9796339 and DMR-0071668.


\begin{thebibliography}{}
%
\bibitem{Feng_PRL}
X. G. Feng, D. Popovi\'c, S. Washburn, and V. Dobrosavljevi\'c, submitted 
to Phys. Rev. Lett.; X. G. Feng, D. Popovi\'c, and S. Washburn, Phys. Rev.
Lett. {\bf 83}, (1999) 368.
%Author, Journal \textbf{Volume,} (year) page numbers.
\bibitem{Krav} S. V. Kravchenko, {\it et al.}, Phys. Rev. B~{\bf 50}, (1994)
8039; S. V. Kravchenko, {\it et al.}, Phys. Rev. B~{\bf 51}, (1995) 7038.
\bibitem{AFS} T. Ando, A. B. Fowler, and F. Stern, Rev. Mod. Phys.~{\bf 54}, 
(1982) 437, and references therein.
\bibitem{Sjostrand} M. E. Sj\"ostrand and P. J. Stiles, Solid State Commun.
{\bf 16}, (1975) 903; M. E. Sj\"ostrand, T. Cole, and P. J. Stiles, Surf. Sci.
{\bf 58}, (1976) 72.
\bibitem{kirstin} K. Walther, D. Popovi\'c, and P. J. Stiles (unpublished).
\bibitem{DP_PRL} D. Popovi\'c, A. B. Fowler, and S. Washburn, Phys. Rev. Lett.
{\bf 79}, (1997) 1543.
\bibitem{Eng} K. Eng, X. G. Feng, D. Popovi\'c, and S. Washburn (to be 
published).
\bibitem{Yoon} J. Yoon, {\it et al.}, Phys. Rev. Lett. {\bf 84}, (2000) 4421.
\bibitem{Sarma} S. Das Sarma and E. H. Hwang, Phys. Rev. Lett. {\bf 84}, (2000)
5596.
\bibitem{par_B} D. Simonian, {\it et al.}, Phys. Rev. Lett. {\bf 79}, (1997)
2304; V. M. Pudalov, {\it et al.}, JETP Lett. {\bf 65}, (1997) 932; S. V.
Kravchenko, {\it et al.}, Phys. Rev. B {\bf 58}, (1998) 3553; D. Simonian, {\it
et al.}, Physica B {\bf 256-258}, (1998) 607; K. M. Mertes, {\it et al.}, 
Phys. Rev. B {\bf 60}, (1999) R5093.
\bibitem{Okamoto} T. Okamoto, {\it et al.}, Phys. Rev. Lett. {\bf 82}, (1999) 
3875.
\bibitem{Kunze} U. Kunze, Phys. Rev. B {\bf 35}, (1987) 9168.
\bibitem{Vitkalov} S. A. Vitkalov, {\it et al.}, Phys. Rev. Lett. {\bf 85},
(2000) 2164.
\bibitem{Pudalov_loc}  V. M. Pudalov, {\it et al.}, preprint cond-mat/0004206
(2000).

\end{thebibliography}
\end{document}